\documentclass[conference]{IEEEtran}
\IEEEoverridecommandlockouts
\usepackage{cite}
\usepackage{amsmath,amssymb,amsfonts}
\usepackage{algorithmic}
\usepackage{graphicx}
\usepackage{textcomp}
\usepackage{xcolor}
\def\BibTeX{{\rm B\kern-.05em{\sc i\kern-.025em b}\kern-.08em
    T\kern-.1667em\lower.7ex\hbox{E}\kern-.125emX}}

\begin{document}

\title{A Reusable Framework Based on Reinforcement Learning to Design Antennas for Curved Surfaces}

\makeatletter
\newcommand{\linebreakand}{%
  \end{@IEEEauthorhalign}
  \hfill\mbox{}\par
  \mbox{}\hfill\begin{@IEEEauthorhalign}
}
\makeatother

\author{\IEEEauthorblockN{Enrique M. Lizarraga$^{1,2}$, Walter E. Herrera$^{2}$}  
\IEEEauthorblockA{
$^{1}$National University of Cordoba, Argentina \\
$^{2}$National University of Catamarca, Argentina\\
Email: emlizarraga@unc.edu.ar; wherrera@tecno.unca.edu.ar}}

\maketitle

\begin{abstract}
The design and implementation of low-profile antennas has been analyzed in past decades from different perspectives while the purpose is to have a small size in the device, and an adequate electromagnetic behavior. This work pursues a methodology to identify small antennas and consequently presents some similarities. Meanwhile, curved surfaces are considered for a certain variety of antennas with reduced size. The so-called \emph{deep reinforcement learning} technique is used as an assistance against morphological variations that are specifically taken into account in this work. 
The objective is to identify antennas that can be efficiently mounted on the surface of metal tubes
such as those frequently present in public infrastructure (e.g. traffic lights and luminaires). 
The motivation is to reduce the visual impact and optimize the radiation pattern of the antenna.
It is analyzed that if changes in variables such as the radius of curvature, or the electromagnetic properties of the materials appear, 
an automatic identification of the underlying characteristics of the problem (by means of \emph{machine learning} techniques)
can readjust the design efficently. 
The results obtained in this work are analyzed based on variables that are typically used to characterize antennas, such as their impedance and radiation pattern.
\end{abstract}


\section{Introduction}

As the development of technologies for wireless communications progresses, there is a marked increase in the density of devices of this nature per unit of area. In this work we concentrate on the antenna systems that are a fundamental part of these devices.
It is then clearly identifiable that the antenna installation density grows in a very significant way.
Given this effect, the need for two actions is highlighted. 

On the one hand, it will be necessary to optimize the radiation pattern of each installation. This principle that seems totally intuitive has an importance that grows exponentially in the face of the need to achieve adequate \emph{electromagnetic compatibility}.
Currently, the selection of an antenna model for any communication device seems to obey directly the principle of optimizing the power budget.
Information for end users, or engineering teams, regarding how to optimize the distribution of electromagnetic energy is frequently omitted in the commercial specification of both, complete communication devices, and elemental antennas.
As the so-called \emph{quality of service} in communication applications is observed degraded in specific uses, the need to take concrete actions in this field is verified.
These days the communications infrastructure for connectivity on public roads is not fully, or significantly, optimized in the sense of electromagnetic compatibility.
The IEEE 802.11 standard is taken as an example since it has massive use and suffers this effect significantly~\cite{interfwifi}.

On the other hand, 
the challenge of 
reducing
the visual impact produced by antenna systems is also highlighted.
Obviously this objective will have an immediate benefit.
The 
ideas presented in this work are delimited to the case of installing antennas on curved surfaces such as those presented by the tubular components of luminaire pole devices or traffic lights. Beyond this example, these ideas can be generalized in various ways as will be presented subsequently.

In recent times, numerous ways have been reported to take advantage of \emph{machine learning} to provide automation or assistance in antenna design. For example, the use of specific techniques such as \emph{simulated annealing}~\cite{RICHIE2017249} or \emph{genetic algorithms} are used to produce specific solutions.
Specific algorithms for satellite antennas have been reported~\cite{antennaRL} where \emph{reinforcement learning} (RL), a technique which seems to be surprisginly effective in many areas, is used.

To add a bit of precision, it should be considered that RL consists of a learning principle based on the \emph{trial and error} mechanism, studied for several decades, through tabular interpretations~\cite{Sutton1998, RPIC19}. But with the advent of a more mature development of the so-called \emph{neural networks}, their interaction with the RL paradigm has been optimized to give rise to what is often called \emph{deep reinforcement learning}~\cite{Wang17_DL}.
The depth suggested by the name refers to the fact that the use of \emph{multilayer neural networks} tends to identify particularities of a certain problem (an optimization problem) that may be imperceptible to the human being that interacts with the algorithm. That is, the human being leaves degrees of freedom in certain variables (neurons) of a neural network and in other variables of the algorithm as a whole, and by means of successive tests, RL optimizes the use of those variables, imitating \emph{learning} of intrinsic properties of the problem.

In this work, these principles are used to establish a framework as follows. A typical antenna is defined, which has gemoetric properties adequate for curved surfaces. Note that the operation of an antenna is dominated by the adjustment of its physical dimensions. An initial crude definition of dimensions is done. Meanwhile, maximum and minimum values for each dimension are set, with the perspective of having antennas with reduced size.
The algorithm is then run cyclically to identify a set of dimension specifications that matches a suitable antenna, in terms of its electromagnetic behavior. As a result we found that the algorithm effectively identifies suitable solutions and also becomes robust to obtain equivalent new antennas for new scenarios, for example, when a different radius of curvature is used.
This observation suggests that a beneficial design methodology is obtained.

This work is organized as follows. 
In Section~\ref{model} the analyzed model is carefully described. 
Details of implementation are presented in Section~\ref{imple}. 
Section~\ref{resul} shows simulation results, 
while conclusions are presented in Section~\ref{conclu}.

\begin{figure}[!t]
\centering
\includegraphics[width=\columnwidth]{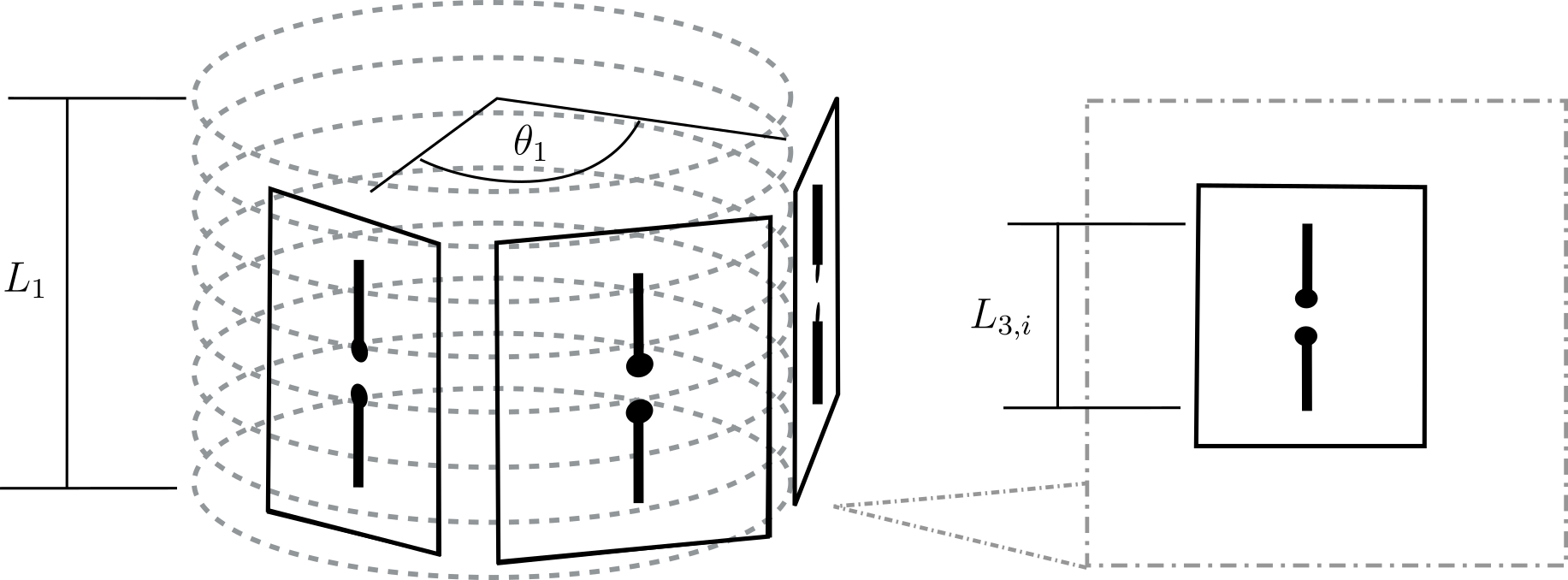}
\caption{Proposed antenna for curved surfaces} 
\label{fig1}
\end{figure}

\section{System Model}
\label{model}

The description of the system model is divided into two parts. On the one hand, the geometric model of the antenna that is considered for this work is indicated. Then 
particular definitions regarding the use of RL to solve the proposed problem are described.

\subsection{Antenna Shape}

A quasi-curved shape has been defined, which is made up of three small planes, as illustrated in the Fig.~\ref{fig1}. In each plane the installation of a small dipole is foreseen. 
The antenna is assumed to be attached to a tube with radius $R1$, from which a fragment of length $L_1$ is modeled. Variable $L1$ tries to avoid the introduction of parts of the tube that does not affect sensitively the electromagnetic behavior in the associated simulation described later.
It is assumed that the length of each dipole is defined by our algorithm, by means of the variables 
$L_{3,i}$ for $i=0,1,2$.
The three small planes previously mentioned are distributed over the tube encompassing an arc given by angle $\theta_1$. This angle can also be defined automatically by the algorithm.
The dipoles are connected in parallel by small transmission lines with a $180^{\text{o}}$ of phase change.
In turn, the distance between each dipole and the conductive tube (i.e. $D1$) is adjusted automatically.
Note that mechanical clamping of antenna components can be achieved on a large scale with the manufacture of suitable plastic parts. However, on a small scale, if a few units are considered, it would be reasonable to assume the manufacture of plastic parts from 3D prints. In this way, the adjustable variables of the antenna geometric model can be precisely manipulated.

\subsection{Problem Modeling with RL}

It is well known that the use of RL requires the definition of an \emph{agent} and an \emph{environment}, as well as a way to characterize the \emph{state} of the environment. In parallel, the possibility of executing \emph{actions} to observe eventual \emph{rewards} is defined~\cite{Sutton1998}. It is frequently mentioned the handling of an \emph{state-action-reward} tuple that defines signals to relate the agent with the environment. The problem is interpreted in this work as follows. The agent is an entity with deep reinforcement learning that intends to learn \emph{how to design antennas for curved surfaces}. It is implemented with a two-layer neural network devoted to implement $Q$-learning~\cite{Sutton1998}. The environment is a computer program based on the \emph{numerical electromagnetic code} specification (NEC2)~\cite{necref} that analyzes the electromagnetic behavior of the antenna. The connection between the agent and the environment (which in turn includes the NEC simulation) is presented in Fig.~\ref{fig1_diagram}.
A simple block stores the current definitions regarding the dimensions of the antenna, and applies the modifications indicated by the action signal. These dimensions are taken as input to define a NEC antenna model in a conventional way.
The output of the NEC simulation is an estimation of the antenna impedance, and a vector representation of the radiation pattern.
\begin{figure}[!t]
\centering
\includegraphics[width=.99\columnwidth]{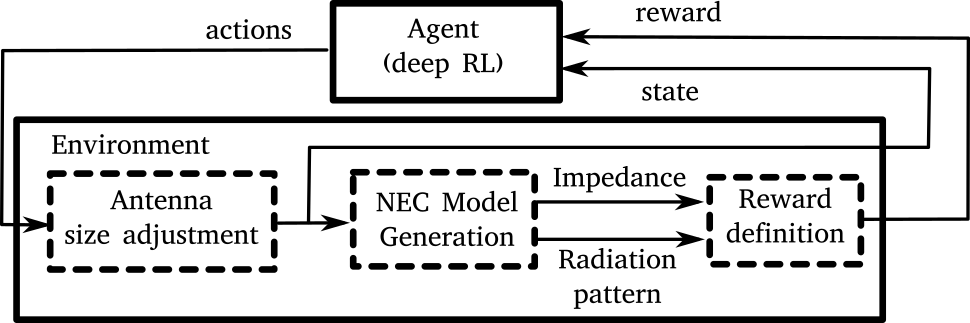}
\caption{Reinforcement learning interpretation of the antenna definition problem.} 
\label{fig1_diagram}
\end{figure}

The state of the system is characterized by the variables: $\theta_1$, the set $\{L_{3,i}\}$ and $D1$.
Note that the simulated antenna is defined according to a certain set of variables that is iteratively adjusted.
A special set of variables is added to the definition of the state, they are simple flags to define whether any value in $\{L_{3,i}\},$ $\theta_1$ or $D1$ have reached a maximum or a minimum. 
These extreme values are set arbitrarily, but practical dimensions are considered for the realistic implementation of the obtained antenna. For example, $D1$ cannot take too high a value, since in that case the small dipoles would move away from the tube and become more visible.
Also, $D1$ cannot be less than a few millimeters,
to allow a simple manufacture of the mechanical support of each dipole, 
and to avoid significant capacitance.

The agent decides (or selects) actions that are applied to the environment on an iterative basis. Actions can be one of these: 
increase the value of $D1$, decrease the value of $D1$, 
increase the value of $\theta_1$, decrease the value of $\theta_1$, increase any of the three values in $\{L_{3,i}\}$, 
decrease any of the three values in $\{L_{3,i}\}$, 
or
do not modify any variable (no operation). 
Increasing and decreasing are specified in steps given by fixed values.
A special signal, the \emph{reward} signal as commonly named in RL~\cite{Sutton1998}, is always set to $-1$, except when the following empirically established condition is reached:
\begin{equation}
	(VSWR<2, G>10[\text{dB}], \varphi<15^{\text{o}}),
\label{eq1}    
\end{equation}
where $\varphi$ is the angle observed for the impedance of the antenna,
$G$ is the gain difference between two normal directions, and $VSWR$ is defined classically~\cite{davidson_2010}.
When the condition in \eqref{eq1} is reached,
a positive value is arbitrarily given by means of the reward signal.
Note that these variables are iteratively redefined, while the algorithm seeks to achieve the maximum reward.

The tube is implemented through a wire mesh and the equal area rule is used to analyze the length and radius of the segments~\cite{davidson_2010}. This assumes a conductive metal tube.

\section{Implementation}
\label{imple}

The main simulation engine for electromagnetic behavior is NEC2. This algorithm has some limitations, although it operates relatively fast and allows us to validate the working hypothesis about the interaction between this problem and the properties of RL.
Two versions of NEC2 are used in turn. On the one hand, PyNEC~\cite{pynec} is used to run a simulation code on certain specific operating systems. Then, taking into account the principle of contrasting results through multiple software packages, 4NEC2~\cite{4nec2} is used to repeat the reports. It is highlighted that 4NEC2 presents a graphical interface that facilitates the presentation of the information.

\section{Results}
\label{resul}

By means of numerical simulation, a configuration such as the one presented in Table~\ref{table_sim_parameters} is analyzed. In our setting, deep \emph{Q-learning}~\cite{Sutton1998} with experience replay and ADAM optimization~\cite{GoodBengCour16} is specifically used.
Meanwhile, Table~\ref{table_sim_parameters_2} shows other settings for the algorithm, in order to ensure a reduced visual impact of the antenna (reduced size), and to allow the algorithm to efficently explore possible solutions.
\begin{table}[!t]
\renewcommand{\arraystretch}{1.3}
\caption{RL Parameters}
\label{table_sim_parameters}
\centering
\begin{tabular}{|c||c|}
\hline
First layer coeffs. count & $100$ \\
\hline
Second layer coeffs. count & $100$ \\
\hline
$\varepsilon$-greedy value for Q-learning & $0.5$ to $0$ \\
\hline
Learn rate $\alpha$ in Q-learning & $0.5$ \\
\hline
Learn rate in ADAM optimization & $10^{-4}$ \\
\hline
\end{tabular}
\end{table}
\begin{table}[!t]
\renewcommand{\arraystretch}{1.3}
\caption{Allowed Values for Antenna Size}
\label{table_sim_parameters_2}
\centering
\begin{tabular}{|c||c|}
\hline
Tube radius $R1$ & $10~[\text{cm}]$\\
\hline
Tube length $L1$ & $25~[\text{cm}]$\\
\hline
Maximum allowed value in $D1$ & $5~[\text{cm}]$\\
\hline
Minimum allowed value in $D1$ & $1~[\text{cm}]$\\
\hline
Fix step in $D1$ adjustement & $1~[\text{cm}]$\\
\hline
Maximum allowed value in each dipole $\{L_{3,i}\}$ & $15~[\text{cm}]$\\
\hline
Minimum allowed value in each dipole of $\{L_{3,i}\}$ & $1~[\text{cm}]$\\
\hline
Fix step in adjustement of each dipole of $\{L_{3,i}\}$ & $5~[\text{mm}]$\\
\hline
Maximum allowed value in $\theta_1$ & $90^{\text{o}}$\\
\hline
Minimum allowed value in $\theta_1$ & $10^{\text{o}}$\\
\hline
Fix step increment in $\theta_1$ adjustement & $1^{\text{o}}$\\
\hline
Fix step decrement in $\theta_1$ adjustement & $0,286^{\text{o}}$\\
\hline
\end{tabular}
\end{table}
Intensive runs of the algorithm have been carried out, which resulted in an \emph{episodic} approach~\cite{Sutton1998}. The following results describe one of the antennas that the algorithm has identified as a valid solution, and is analyzed to show the flexibility of this strategy. 
The antenna described below results from the application of the proposed algorithm, and has the characteristics shown in Table~\ref{table_antenna}. Note that in successive iterations, on a random basis, the algorithm can deliver new antennas with new characteristics, but each solution will respect the specifications in~\eqref{eq1}. Therefore, the antenna specifications discussed in the following do not represent a deterministic output of the algorithm, but rather belong to a set of solutions that the algorithm efficiently produces.
In Fig.~\ref{fig3} the obtained radiation pattern for the antenna specified in Table~\ref{table_antenna} is presented. The radial distribution of energy is confined in a certain area. 
This representation allows us to 
propose
that the obtained antenna is suitable to be installed on the edges of a region that is to be served with connectivity, without excessively radiating in directions that do not correspond to the addressed service.
Figure~\ref{fig4} shows the excellent $VSWR$ that has been achieved with respect to a $50$~$\Omega$ transmission line.
The frequency response is then analyzed by means of Fig.~\ref{fig5}, and the Fig.~\ref{fig6} is presented as a validation
of the geometry that is effectively simulated.
Note that Fig.~\ref{fig6} complements Fig.~\ref{fig3}, and brings more details regarding the radiation pattern.

\begin{figure}[!t]
\centering
\includegraphics[width=\columnwidth]{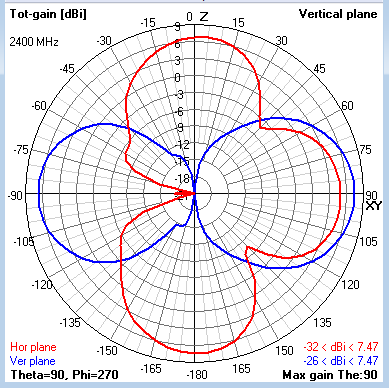}
\caption{Radiation pattern.} 
\label{fig3}
\end{figure}

\begin{figure}[!t]
\centering
\includegraphics[width=.4\columnwidth]{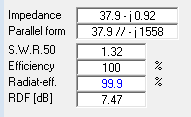}
\caption{Impedance simulated in software \texttt{4nec2}.} 
\label{fig4}
\end{figure}

\begin{figure}[!t]
\centering
\includegraphics[width=\columnwidth]{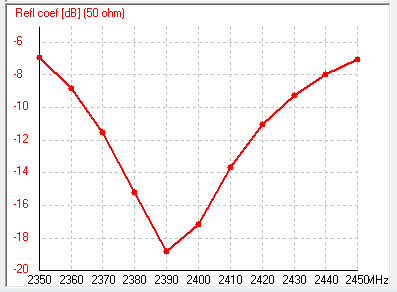}
\caption{Frequency response.} 
\label{fig5}
\end{figure}


\begin{figure*}[!t]
\centerline{
a){\includegraphics[width=.7\columnwidth]{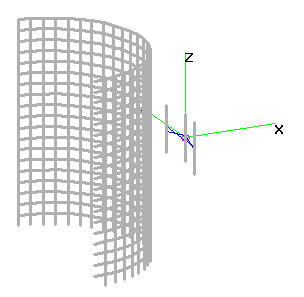}
\label{fig6a}}
\hfil
b){\includegraphics[width=.9\columnwidth]{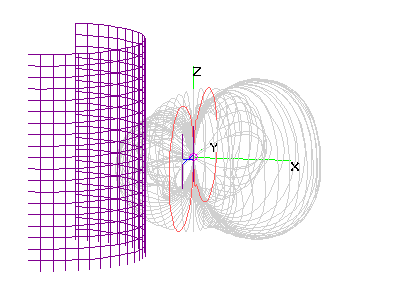}
\label{fig6b}}
}
\caption{A view of an obtained antenna. a) Representation of the metallic components that are simulated. The axes indicate $0.1$~[m] scale. b) 3D rendering of the radiation pattern.}
\label{fig6}
\end{figure*}

The possibility of reusing the learning that our algorithm iteratively acquires has been studied, by means of the introduction of new specifications that were not used to generate the original learning. The idea behind this experiment is to evaluate the capacity of the proposed methodology to generalize the calculation procedure, automatically.
When referencing new specifications, a change in the tube radius can be taken into account as the most expected case. However, it is noted that even an improvement in the electromagnetic simulation tool (replacing NEC2) belongs to this range of possibilities. 

In this sense, it would be possible to inquire about the use of \emph{a single neural network instead of RL}. In principle, this setting could implement a transformation relationship between the specifications of the antenna (for example, radius of curvature of the tube) and the dimensions that give an optimal antenna. This perspective has been studied previously by other authors and with different antenna topologies, by using a big number of antenna simulations to train the network, which can introduce  a huge processing load. However, in this work an algorithm is intended that, in the face of variations in the specifications, is capable of obtaining an adequate antenna by reusing the information acquired from a previous adjustment problem, efficiently. In the literature there is evidence that this perspective is not naturally embedded when a neural network is used directly. In the case of RL, this characteristic is indeed introduced. By way of comparison, we can cite the case where neural networks are used to identify handwritten digits, and a large number of manuscripts is required to adjust the network. But RL is used to teach a robot to take steps, in a unique setting (in a single floor or room). The learning achieved in that experiment (the walking in specific conditions) is then applied to many surfaces (i.e. different types of floor, different friction, different weight, etc.). 

\begin{table}[!t]
\renewcommand{\arraystretch}{1.3}
\caption{Obtained antenna for $R1=0,1~[m]$}
\label{table_antenna}
\centering
\begin{tabular}{|c||c|}
\hline
Separation $D1$ & $5~[\text{cm}]$\\
\hline
First dipole length $\{L_{3,0}\}$ & $6~[\text{cm}]$\\
\hline
Second dipole length $\{L_{3,1}\}$ & $5,5~[\text{cm}]$\\
\hline
Third dipole length $\{L_{3,2}\}$ & $6~[\text{cm}]$\\
\hline
Angular separation $\theta_1$ & $25,4^{\text{o}}$\\
\hline
\end{tabular}
\end{table}

\begin{table}[!t]
\renewcommand{\arraystretch}{1.3}
\caption{Obtained antenna for $R1=0,12~[m]$}
\label{table_antenna_assessment}
\centering
\begin{tabular}{|c||c|}
\hline
Separation $D1$ & $5~[\text{cm}]$\\
\hline
First dipole length $\{L_{3,0}\}$ & $6~[\text{cm}]$\\
\hline
Second dipole length $\{L_{3,1}\}$ & $5,5~[\text{cm}]$\\
\hline
Third dipole length $\{L_{3,2}\}$ & $5~[\text{cm}]$\\
\hline
Angular separation $\theta_1$ & $24,99^{\text{o}}$\\
\hline
\end{tabular}
\end{table}

For the algorithm treated here, the number of attempts it takes to identify a new antenna according to~\eqref{eq1} has been observed from this perspective. If any variable changes in the problem definition (for example, by varying the radius of curvature $R1$), it has been observed that the training previously elaborated according to Table~\ref{table_sim_parameters_2} generates a good speed of adaptation to new cases, which suggest efficiency.
Then, an analysis of the efficiency of the algorithm is carried out using two perspectives, which are made up of three experiments.
On the one hand, a \emph{learning stage} (training) for the algorithm is defined. 
This stage is what is considered a first experiment. In this stage, $20,000$ specific antenna simulations are run, which are determined by the iterative execution of the \emph{deep reinforcement learning} paradigm embedded in our algorithm. In this stage the implementation of the \emph{trial and error} principle is executed. During this cycle, the number of attempts that are performed until 
an
adequate antenna appears, meeting the specification in~\eqref{eq1}, is observed. This result is presented in Fig.~\ref{fig7}. Based on this it is interpreted that the algorithm effectively identifies underlying characteristics in the solution of the problem (adjustment of the dimensions of the antenna) and thus provides solutions with increasingly frequent success, as the algorithm evolves.
This is, vertical values in Fig.~\ref{fig7} decreases as the right part of the horizontal axis is considered. This is taken as a first verification of its efficiency.
On the other hand, efficiency is studied from a different point of view. The algorithm is run a second time (second experiment) with a number of iterations limited to 1250 runs. However, a new radius of curvature (i.e. $R1=0,12$~[m]) is now defined for the tube. Note that this means significantly modifying the problem to be solved. In this condition, the training executed in the first experiment of the algorithm is not recovered (not acquired), so the algorithm starts without any experience with the problem. In this case, the algorithm identifies different solutions, from which one in particular with $VSWR = 1.23$ is extracted, which results in the best solution found. A third experiment of the algorithm is finally performed. The algorithm is restarted, although in this case the previously saved state for the neural network is acquired, which implies incorporating the previous experience acquired during the first experiment. In this case, the algorithm identifies different solutions, from which one in particular with $VSWR = 1.13$ is extracted, which is the best solution found. This solution is presented in the 
Table~\ref{table_antenna_assessment}.
Note that the best performance of this antenna (which belongs to a new design problem, which is not the one used for training) suggests that the framework proposed by this work is reusable and efficient. 
Given these observations, it is expected to obtain improved results by replacing the small dipoles described previously  by other variants (such as patches). However, the presented simulation results objectively show that the proposed framework responds satisfactorily.



\begin{figure}[!t]
\centering
\includegraphics[width=.6\columnwidth]{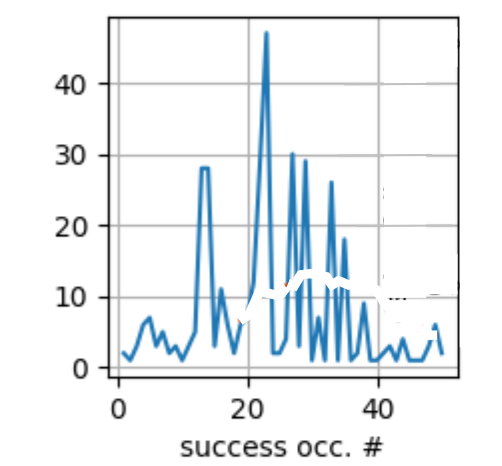}
\caption{Evolution of the algorithm. On the horizontal axis the number of identified antenna is represented (meeting the specifications), on the vertical axis the number of attempts that elapsed until each antenna was identified is indicated.} 
\label{fig7}
\end{figure}

\section{Conclusions}
\label{conclu}

A strategy has been presented which defines a specific methodology to design curved antennas that are applied on conductive material tubes. Reinforcement learning is used to automatically incorporate underlying characteristics of the problem, without having to explicitly model them. 
The algorithm reuses  information acquired in the learning stage, to adequately resolve the antenna sizing under different problem settings. The general typology of antennas supported by this framework has been empirically defined, and can be enriched with additional details. However it responds adequately, taking into consideration analysis values such as the radiation pattern and impedance, as well as the frequency response.
The existence of antennas that tend to reduce the visual impact of the infrastructure of the wireless communications system, and to favor electromagnetic compatibility, is verified. 
\nocite{fig6a}

\bibliographystyle{IEEEtran}
\bibliography{IEEEabrv,IEEECONF_ARGENCON20_refs}

\end{document}